\documentclass[fleqn,twoside]{article}
\usepackage{espcrc2}
\usepackage{graphicx}
\usepackage[figuresright]{rotating}

\RequirePackage{xspace}

\usepackage{relsize}
\def\babar{\mbox{\slshape B\kern-0.1em{\smaller A}\kern-0.1em
    B\kern-0.1em{\smaller A\kern-0.2em R}}}

\def\Kbar  {\kern 0.2em\overline{\kern -0.2em K}{}\xspace}

\def\Kz    {\ensuremath{K^0}\xspace}
\def\Kzb   {\ensuremath{\Kbar^0}\xspace}
\def\KzKzb {\ensuremath{\Kz \kern -0.16em \Kzb}\xspace}
\def\Kp    {\ensuremath{K^+}\xspace}
\def\Km    {\ensuremath{K^-}\xspace}

\def\KpKm  {\ensuremath{\Kp \kern -0.16em \Km}\xspace}

\def\Dbar    {\kern 0.2em\overline{\kern -0.2em D}{}\xspace}

\def\Dz      {\ensuremath{D^0}\xspace}
\def\Dzb     {\ensuremath{\Dbar^0}\xspace}
\def\DzDzb   {\ensuremath{\Dz {\kern -0.16em \Dzb}}\xspace}
\def\Dp      {\ensuremath{D^+}\xspace}
\def\Dm      {\ensuremath{D^-}\xspace}

\def\DpDm    {\ensuremath{\Dp {\kern -0.16em \Dm}}\xspace}

\def\Bbar    {\kern 0.18em\overline{\kern -0.18em B}{}\xspace}

\def\Bz      {\ensuremath{B^0}\xspace}
\def\Bzb     {\ensuremath{\Bbar^0}\xspace}
\def\BzBzb   {\ensuremath{\Bz {\kern -0.16em \Bzb}}\xspace}
\def\Bu      {\ensuremath{B^+}\xspace}
\def\Bub     {\ensuremath{B^-}\xspace}

\def\BpBm    {\ensuremath{\Bu {\kern -0.16em \Bub}}\xspace}

\mathchardef\Upsilon="7107
\def\Y#1S{\ensuremath{\Upsilon{(#1S)}}\xspace}% no space before {...}!

\mathchardef\Deltares="7101
\mathchardef\Xi="7104
\mathchardef\Lambda="7103
\mathchardef\Sigma="7106
\mathchardef\Omega="710A

\def\Deltabar{\kern 0.25em\overline{\kern -0.25em \Deltares}{}\xspace}
\def\Lbar{\kern 0.2em\overline{\kern -0.2em\Lambda\kern 0.05em}\kern-0.05em{}\xspace}
\def\Sigbar{\kern 0.2em\overline{\kern -0.2em \Sigma}{}\xspace}
\def\Xibar{\kern 0.2em\overline{\kern -0.2em \Xi}{}\xspace}
\def\Obar{\kern 0.2em\overline{\kern -0.2em \Omega}{}\xspace}
\def\Nbar{\kern 0.2em\overline{\kern -0.2em N}{}\xspace}
\def\Xb{\kern 0.2em\overline{\kern -0.2em X}{}\xspace}

\newcommand{\tev}{\ensuremath{\mathrm{\,Te\kern -0.1em V}}\xspace}
\newcommand{\gev}{\ensuremath{\mathrm{\,Ge\kern -0.1em V}}\xspace}
\newcommand{\mev}{\ensuremath{\mathrm{\,Me\kern -0.1em V}}\xspace}
\newcommand{\kev}{\ensuremath{\mathrm{\,ke\kern -0.1em V}}\xspace}
\newcommand{\ev}{\ensuremath{\mathrm{\,e\kern -0.1em V}}\xspace}
\newcommand{\gevc}{\ensuremath{{\mathrm{\,Ge\kern -0.1em V\!/}c}}\xspace}
\newcommand{\mevc}{\ensuremath{{\mathrm{\,Me\kern -0.1em V\!/}c}}\xspace}
\newcommand{\gevcc}{\ensuremath{{\mathrm{\,Ge\kern -0.1em V\!/}c^2}}\xspace}
\newcommand{\mevcc}{\ensuremath{{\mathrm{\,Me\kern -0.1em V\!/}c^2}}\xspace}
\def\mus  {\ensuremath{\rm \,\mus}\xspace}

\def\mus        {\ensuremath{\,\mu{\rm s}}\xspace}    %% microsecond
\def\to                 {\ensuremath{\rightarrow}\xspace}

\def\pep2{PEP-II}

\def\PM #1 #2   {^{+#1}_{-#2}{}}

\def\gsim{{~\raise.15em\hbox{$>$}\kern-.85em
          \lower.35em\hbox{$\sim$}~}\xspace}
\def\lsim{{~\raise.15em\hbox{$<$}\kern-.85em
          \lower.35em\hbox{$\sim$}~}\xspace}

\xspace

\newcommand{\epjBase}        {Eur.\ Phys.\ Jour.\xspace}

\newcommand{\jprBase}        {Phys.\ Rev.\xspace}
\newcommand{\jplBase}        {Phys.\ Lett.\xspace}
\newcommand{\nimBaseA}       {Nucl.\ Instr.\ Meth.\xspace}

\newcommand{\nimBaseC}       {Nucl.\ Instr.\ and Methods\xspace}

\newcommand{\npBase}         {Nucl.\ Phys.\xspace}
\newcommand{\zpBase}         {Z.\ Phys.\xspace}

\newcommand{\epjc}      [1]  {\epjBase\ C~{\bf #1}}

\newcommand{\mpl}       [1]  {{Mod.\ Phys.\ Lett.\ {\bf #1}}}

\newcommand{\nim}       [1]  {\nimBaseC~{\bf #1}}
\newcommand{\nima}      [1]  {\nimBaseA~A~{\bf #1}}

\newcommand{\npb}       [1]  {\npBase\ B~{\bf #1}}

\newcommand{\npbps}     [1]  {{Nucl.\ Phys.\ B~Proc.\ Suppl.\ {\bf #1}}}

\newcommand{\plb}       [1]  {\jplBase\ B~{\bf #1}}

\newcommand{\pr}        [1]  {\jprBase\ {\bf #1}}

\newcommand{\progtp}    [1]  {{Prog.\ Th.\ Phys.\ {\bf #1}}}

\newcommand{\zpc}       [1]  {\zpBase\ C~{\bf #1}}

\def\jetset74   {\mbox{\tt Jetset \hspace{-0.5em}7.\hspace{-0.2em}4}\xspace}
 \def\er #1 #2 { $#1 \pm #2$ }
 \def\bra #1 #2 #3 #4 { $#1 ^{+#2} _{-#3} \pm #4 $ }

\title{Long distance effects in semi-inclusive B decays}

\author{Ugo Aglietti \address[MCSD]{
Dipartimento di Fisica, Universit\`a di Roma I ``La Sapienza'',
and I.N.F.N., Sezione di Roma, Italy} and Giulia Ricciardi
\address[MCSD]{Dipartimento di Scienze Fisiche,
Universit\`a di Napoli ``Federico II'' and I.N.F.N., Sezione di
Napoli, Italy.}}

\begin{document}

\begin{abstract}
We discuss some issues on factorization of long distance effects
for semi-inclusive $B$ decay spectra in full QCD and in the
effective theory. \vspace{1pc}
\end{abstract}

\maketitle

\section{Introduction}

Let us consider B meson decays into a QCD jet plus non QCD
partons, such as a real photon or a lepton pair. This decay is
characterized by three fundamental mass scales, the heavy flavor
mass $m_b$ and the energy $E_X$ and invariant mass of the jet
$m_X$. The threshold region is defined as the region where $m_X
\ll E_X$, that is the region where the final observed state has
large energy and the emission of real soft and collinear partons
is inhibited. Close to threshold, in the perturbative calculation,
the partial cancellation of infrared real and virtual
contributions produces  large logarithms in the ratio of $m_X$ and
the hard scale $Q$. In the heavy quark effective theory (HQET),
the heavy quark is replaced by a static color charge, leaving
unchanged infrared exchanges in the heavy meson, and therefore the
structure of the large infrared logarithms. By integrating out the
heavy flavor mass in HQET, the only remaining scales in the
hadronic subprocess are $m_X$ and $E_X$ and the hard scale $Q$ is
identified with $E_X$. Infrared large logarithms at threshold
occur in the ratio $E_X/m_X$ \cite{ugosol,noi1}. Such logarithms
have to be resummed in order not to spoil the perturbative
expansion. We will elaborate on some issues related to the
universality of long distance effects.

\section{Universality of QCD form factor}
\label{sect-univ}

For the semileptonic decay $B \rightarrow X_u l \nu$, the most
general differential distribution can be written in a factorized
form:
\begin{eqnarray}
\frac{1}{\Gamma}\frac{d^3\Gamma}{dx du dw} &=&
C\left[x,w;\alpha(w\,m_b)\right]\,
\sigma\left[u;\alpha(w\,m_b)\right] \, + \nonumber
\\&+&\, d\left[x,u,w;\alpha(w\,m_b)\right]~,
\label{tripla}
\end{eqnarray}
where
\begin{equation}
w \,=\, \frac{2 E_X}{m_b},\qquad x \,=\, \frac{2 E_l}{m_b}
\label{wx}
\end{equation}
and
\begin{equation}
u \, = \, \frac{E_X - \sqrt{E_X^2 - m_X^2} }{E_X + \sqrt{E_X^2 -
m_X^2} } \, \approx \, \frac{m_X^2}{4 \, E_X^2}. \label{u}
\end{equation}
The distribution is normalized to the radiatively-corrected total
semileptonic width $\Gamma $. We have two short-distance,
process-depen\-dent functions,  the coefficient function
$C\left[x,w;\alpha(w\,m_b)\right]$ and the remainder function
$d\left[x,u,w;\alpha(w\,m_b)\right]$. All the infrared logarithms
are resummed into
 a
universal, long-distan\-ce dominated, QCD form factor
$\sigma\left[u;\alpha(w\,m_b)\right]$.
 The running
coupling constant is set to the hard scale of the system, $Q$,
where $Q\,= \,w\,m_b \,=\, 2 E_X$.

In the radiative decay $B \rightarrow X_s \, \gamma$, a similar
formula holds for the single differential spectrum in the hadronic
mass
\begin{eqnarray}
\label{radsum2} \frac{1}{\Gamma_R}\frac{d\Gamma_R}{d t_s} \,&=&\,
C_R\left[\alpha(m_b)\right]\, \sigma\left[t_s;\,\alpha(m_b)\right]
\, + \nonumber\\
&+& \, d_R\left[t_s;\,\alpha(m_b)\right],
\end{eqnarray}
where $t_s \, = \, m_{X_s}^2 /m_b^2 $. The form factor $\sigma$,
being the universal form factor, is the same appearing in formula
(\ref{tripla}), while the coefficient and the remainder function,
dependent on the process, are different. In the radiative decay,
$w\sim 1$,  allowing the replacement $\alpha(w\,m_b) \,
\rightarrow \, \alpha(m_b)$; the distribution (\ref{radsum2})
contains a constant coupling $\alpha(m_b)\simeq 0.22$. Due to
kinematics, such replacement cannot be done in the semileptonic
differential distribution (\ref{tripla}). The important issue is
that the form factor $\sigma$, although being universal, depends
 on one kinematical variable $m_{X_s}^2 /m_b^2 $ in the
radiative case, while it depends on two kinematical variables, $u
$ and $w$, in the semileptonic case. Such difference leads to a
natural division into two classes for the double and single
differential distributions, obtained integrating the general
triple differential distribution (\ref{tripla})
\cite{noi1,noi2,noi3}. The first class contains distributions not
integrated over the energy $E_X$, f.i. the single distribution in
$E_X$: they have the same form factor as radiative decays. The
second class contains distributions integrated over the energy
$E_X$, f.i. the single distribution in $m_X$ or in $E_l$: their
dependence on the universal form factor has been spoiled by the
integration, and they have the same form factor as radiative
decays only at leading logarithmic order. Since soft logarithms
signal long distance effects, there is a different long distance
structure with respect to distributions in the first class and to
radiative decay, due to the integration over the hard scale. In
other terms, while in radiative decays the form factor $\sigma$ is
fixed at a hard scale $Q=m_b$,  in the second class the form
factor is integrated on $Q$ with a generic weight function
$\phi(Q)$, dependent on kinematics, from low values up to $m_b$.

%
%%%%%%%%%%%%%%%%%%%%%%%%%%%%%%%%%%%%%%%%%%%%%%%%%%%%%%%%%%%%%%%%%%%%%%%%%%%%%%%%%%%%
%
\section{Semileptonic spectra}

Let us consider, f.i., the hadron energy spectrum, $ d \, \Gamma
/d\,E_X $, belonging to the first class mentioned before. As well
known, its shape presents the so called Sudakov shoulder, related
to the occurrence of infrared singularities close to the threshold
$E_X=m_b/2$. At lowest fixed order $O(\alpha)$, the parton process
$b \rightarrow u\, l \, \nu \, g$, where $g$ is a real gluon,
contributes to the decay. Above the threshold, at  $E_X=E_u+E_g
\ge m_b/2$, large infrared logarithms, of the form
$\log(E_X-m_b/2)$, appear  and become singular at $E_X \rightarrow
m_b/2^+$. Virtual contributions cannot cancel such singularity,
since in the virtual process $b \rightarrow u \, l \, \nu$ the
energy of the up quark, $E_u=E_X$, cannot exceed the energy
$m_b/2$. The solution is to abandon the fixed order calculation
and resum into a form factor all large infrared logarithms.
Resummation completely eliminates the singularity, leaving only an
effect in the characteristic Sudakov shoulder.

The single differential distribution in $E_X$ is obtained
 by integrating
(\ref{tripla}). Let us use the adimensional variables defined in
(\ref{wx}) and (\ref{u}).
 Since there are large logarithms only for $ w > 1$, we
are interested in the resummed formula  in that region only. We
have \cite{noi1}:
\begin{eqnarray}
\label{wg1impr} \frac{1}{\Gamma}\frac{d\Gamma}{dw} &=&
C_{1}\left(\alpha\right) \Big\{1  -  C_{2} \left(\alpha\right)\,
\Sigma\left[w-1;\,\alpha(m_b)\right] \Big\}
 \nonumber\\
&+&R(w;\,\alpha) ,
\end{eqnarray}
where $C_{1}\left(\alpha\right)$ and $C_{2}\left(\alpha\right)$
are two coefficient functions, and $R(w;\,\alpha)$ is a remainder
function, vanishing at $w=1$. They have a perturbative  expansion
 in $\alpha$, whose coefficients are constant in the case of the coefficients,
 depend on $w$ for the remainder.
  The form factor $\Sigma$ is the partially integrated
or cumulative form factor $\Sigma(u,\alpha)$, defined as:
\begin{equation}
\Sigma(u;\,\alpha) \,=\,\int_0^u d u^\prime \,
\sigma(u^\prime;\,\alpha). \label{cumulative}
\end{equation}
It resums all infrared logarithms, and it is universal, being the
same as in the radiative decay; at any order in perturbation
theory the hadron energy semileptonic spectra and the radiative
one have the same infrared logarithms. The form factor
$\Sigma(u;\,\alpha)$ has an exponential form:
\begin{equation}
\qquad \Sigma \, = \, e^G, \label{valeancora0}
\end{equation}
with
\begin{eqnarray}
G(u;\alpha) \,&=& \, \sum_{n=1}^{\infty} \sum_{k=1}^{n+1} G_{n
k}\, \alpha^n \, L^k, \nonumber \\
L \, &\equiv& \, \log\frac{1}{u}.
\end{eqnarray}

The situation changes radically if we take a differential spectrum
belonging to the second class defined above, f.i. the single
differential semileptonic spectrum in the hadronic mass
 or the distribution in   the light--cone momentum,
normalized to $m_b$, $p_+ = (E_X - |\vec{p}_X|)/m_b $. After
integration, such distributions can be rearranged in a factorized
form, with the requirement that the form factor contains all the
infrared logarithms; however, the logarithmic tower will be
different with respect to the radiative decay \cite{noi2,noi3}.

F.i., in the distribution $1/ \Gamma \, d\Gamma/d {p}_+$, we can
construct a cumulative form factor $\Sigma_P(p_+;\,\alpha)$ in
analogy with (\ref{cumulative}). The related form factor can also
be exponentiated:
\begin{equation}
 \Sigma_P \, = \, e^{ G_P }  \label{valeancora}
\end{equation}
with
\begin{eqnarray}
G_P\left(p_+;\,\alpha\right) \,&=&\, \sum_{n=1}^{\infty}
\sum_{k=1}^{n+1} G_{ P n k}\, \alpha^n \, L_P^k,  \nonumber \\
L_P \, &\equiv& \, \log \frac{1}{p_+}.
\end{eqnarray}

 The difference in soft
logarithms explicitly shows up when we compare the numerical
coefficients  $G_{n k}$ and  $ G_{P n k}$:
\begin{eqnarray}
{G}_{P 12} &=& G_{12}, \nonumber
\\
{G}_{P 11} &=& G_{11} + 0.18, \nonumber
\\
{G}_{P 23} &=& G_{23},\nonumber
\\
{G}_{P 22} &=& G_{22}+ 0.089-0.0046 \, n_f,\nonumber
\end{eqnarray}
and so on; $n_f$ is the number of active flavors.

\section{Soft and collinear emissions}

The QCD form factor $\Sigma$ can be written as a convolution of
two functions, each of them corresponding to distinct sources of
large corrections: the shape function $f$, resumming infrared
logarithms due to  soft contributions, and a coefficient or jet
function $J$. The shape function has the characteristics of a
quark distribution function and it is well defined in HQET.
 Upon introducing a factorization scale $\mu_F$ we
can write
\begin{equation}
\Sigma[u;\,\alpha(Q)]= J(u; Q, \mu_F)\;  \otimes \; f(u; \mu_F).
\end{equation}
The coefficient function resums all hard collinear logarithms, as
well as soft effects between the factorization scale $\mu_F$ and
the hard scale of QCD. The factorization scale $\mu_F$ can be
identified with the ultraviolet cut off of the shape function in
the HQET.  The physical basis of such factorization is that soft
logarithms are related to longer distance effects with respect to
hard collinear logarithms. In the radiative case $Q=m_b$, and it
is a natural choice setting $\mu_F=m_b$ as well:
\begin{equation}
\Sigma(m_b)= J(m_b)\; \; f(m_b).
\end{equation}
In the semileptonic case, that is not always the case. At the end
of  section (\ref{sect-univ}) we have represented schematically
the construction of single and double differential decay spectra
as an integration in the hard scale $Q$ weighted by a generic
function
 $\phi(Q)$, dependent on kinematics. That simplification is helpful
 here as well. Let us write
 \begin{eqnarray}
  & & \int_0^{m_b} dQ \, \phi(Q) \; \sigma(Q) =\nonumber \\
 &=& \int_0^{m_b} dQ \, \phi(Q) \; J(Q, \mu_F)\;  f(\mu_F).
 \end{eqnarray}
In order to avoid  substantial soft effects in the jet factor, one
has to take $\mu_F$ of order $Q$. Since Q is integrated over,
 also $\mu_F$ has to be changed and one has to know the
shape function as a function of $\mu_F$. This is to be contrasted
with the radiative case, in which it is sufficient to know the
shape function at the single point $\mu_F=m_b$. Let us stress that
it is essential to factorize all the soft logarithms in the shape
function. One can then replace the perturbatively evaluation of
the shape function with a non perturbative one (f.i. lattice QCD),
including also non perturbative soft effects, such as Fermi motion
of the $b$ quark inside the $B$ meson.

\section{Long distance effects}

Quite often, non perturbative long distance effects are  included
in a phenomenological way  by convoluting the perturbatively
calculated spectrum with a non perturbative structure function.
Being non perturbative, the latter is generally not computed, but,
f.i., parameterized and fitted to the experimental spectrum in the
radiative decay and then used into the theoretical prediction for
the semileptonic decay. This procedure is not without risks,
since, as seen before, the correlation between different decays is
not always immediate, even at the perturbative level. It
introduces the dominant source of uncertainty in the CKM
determination of $V_{ub}$ from the semileptonic decay.

Another possibility is to include such effects without convoluting
the perturbative spectrum with a non perturbative structure
function, thus remaining within the perturbation theory framework.
Effects due to the motion of the $b$ quark inside the meson will
be taken into account  modifying the QCD running coupling constant
into an effective coupling according to a model \cite{noi4}. The
effective coupling will also help solving regularization problems
present already in the perturbation theory, that are traditionally
cured by means of arbitrary prescriptions. They are due to the
fact that resummation formulas do not exclude very low energy
kinematical regions, where the running coupling constant  hits the
Landau pole. The effective coupling is built in two steps
\cite{noi4}. First,
 a dispersion relation is used to analytically extend the running
 coupling constant $\alpha$ to a  coupling $\bar
 \alpha $ without the Landau pole; $\bar
 \alpha $ has the same physical discontinuities than $\alpha$ and
 the same high energy behavior, but it has a finite limit at zero
momentum transfer. The second step consists of resumming in the
effective coupling $\tilde{\alpha}$ the secondary emissions off
the radiated gluons. The resummation formula is the standard one,
but it is controlled by the discontinuity of $\bar
 \alpha $, instead than $\alpha$:
 \begin{equation}
\label{deftime2} \tilde{\alpha}(k_{\perp}^2) \, = \, \frac{i}{2
\pi} \, \int_0^{k_{\perp}^2} \, d s \, {\rm Disc}_s \, \frac{
\bar{\alpha}(- s) }{ s }.
\end{equation}
The effective coupling has an expansion in powers of $\alpha$ and
its usage  can be considered a change of scheme. As in the
previous sections, we can factorize the decay spectra and
calculate the universal form factor $\sigma$. The difference is
that now in the perturbative calculation the effective coupling
 is used. Factorization is easily performed in the Mellin space:
 \begin{equation}
\sigma_N(\alpha) \, = \, \int_0^1 \, (1-u)^{N-1} \,
\sigma(u;\,\alpha) \, du.
\end{equation}
The form factor has an exponential form in $N$-space:
\begin{equation}
\sigma_N(\alpha) \, = \, e^{ G_N(\alpha) },
\end{equation}
where the exponent of the form factor reads:
\begin{eqnarray}
\label{expoGN} & & G_N(\alpha) \, = \, \int_0^1 \frac{dy}{y}
\left[ (1-y)^{N-1} - 1 \right] \nonumber \\  & &\left\{  \int_{Q^2
y^2}^{Q^2 y} \frac{dk_{\perp}^2}{k_{\perp}^2}
\tilde{A}\left[\tilde{\alpha}(k_{\perp}^2)\right] \, + \,
\tilde{B}\left[\tilde{\alpha}(Q^2 y)\right] \right. \nonumber \\
& &  \left. + \, \tilde{D}\left[\tilde{\alpha}(Q^2 y^2)\right]
\right\}.
\end{eqnarray}
This formula is the same as the standard resummation formula,
where effective couplings and functions have replaced  the
standard ones. The functions $\tilde{A}(\tilde{\alpha}),\,
\tilde{B}(\tilde{\alpha})$ and $\tilde{D}(\tilde{\alpha})$ have
expansions in powers of the effective coupling, and are obtained
by matching order for order in $\alpha$ the standard formula with
(\ref{expoGN}).

The form factor in momentum space is obtained by inverse
transform:
\begin{equation}
\label{inverse} \sigma(t; \, \alpha) \, = \,
\int_{C-i\infty}^{C+i\infty} \frac{dN}{2\pi i} (1-t)^{-N} \,
\sigma_N(\alpha),
\end{equation}
where the constant $C$ is chosen so that the integration contour
in the $N$-plane lies to the right of all the singularities of
$\sigma_N(\alpha)$. As anticipated, no prescription is needed
because $\sigma_N(\alpha)$ is analytic for ${\rm Re} \, N \, > \,
0$.

\section{Comparison with data}

\begin{figure}[htbp]
\begin{center}
 \begin{tabular}{c}
   \mbox{\includegraphics[width=7cm,height=5cm]{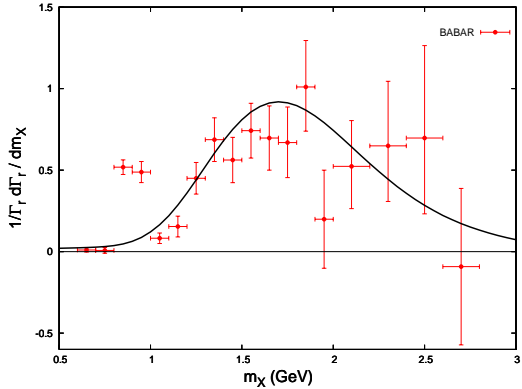}}
 \end{tabular}
\end{center}
\caption{\small{Comparison between theory and Babar data for the
$B\to X_s \gamma$ invariant hadron mass distribution at
$\alpha(m_Z)\, = \, 0.123$.}}
\label{rdmx}
\end{figure}

In Fig.~\ref{rdmx}  the invariant hadron mass distribution for the
radiative decay, $d\Gamma_r/d m_X$ is compared with experimental
data from the BaBar collaboration \cite{babargam2}. The
theoretical curve, calculated in the model described in the
previous section \cite{noi4}, shows a good agreement with data.
The model has no free parameters; however, one can choose to best
fit the theoretical curve to data by using as free parameter
$\alpha(m_Z)$, and obtain $\alpha_S(m_Z) \, = \, 0.123 \, \pm \,
0.003 $.  Let us observe that the data with $m_X < 1.1 $ GeV are
not representative; they show the $K^\star$ peak, which cannot be
reproduced in this approach.

\begin{figure}[htbp]
\begin{center}
 \begin{tabular}{c}
   \mbox{\includegraphics[width=7cm,height=5cm]{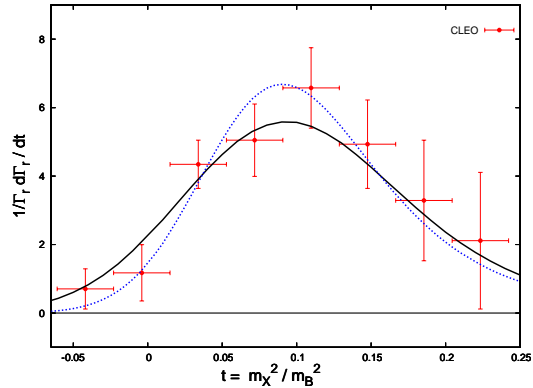}}
 \end{tabular}
\end{center}
\caption{\small{Comparison between theory and Cleo data for $B\to
X_s \gamma$ photon spectrum. Dotted line (blue): $\alpha_S(m_Z) \,
= \, 0.118$ and $\sigma_\gamma \, = \, 100$ MeV to model the
Doppler effect (see text); continuous line (black): $\alpha_S(m_Z)
\, = \, 0.117$ and $\sigma_\gamma \, = \, 150$ MeV.}}
\label{efcleo}
\end{figure}

Another good agreement is obtained by comparing the photon energy
spectrum with data from the Cleo Collaboration \cite{cleogam1}, as
shown in Fig.~\ref{efcleo}. In the $B$ rest-frame, $ t \, = \,
m_X^2/m_B^2 \, = \,  1 \, - \, 2E_{\gamma}/m_B$. The photon
energies are however measured in the $\Upsilon(4S)$ rest frame, in
which the $B$ mesons have a small, non-relativistic motion. In
order to model the Doppler effect, it is possible to convolute the
theoretical curve for $E_\gamma$ --- computed with a $B$ meson at
rest --- with a normal distribution of $\sigma_\gamma \, = \, 100$
or $150$  MeV.

\begin{figure}[htbp]
\begin{center}
 \begin{tabular}{c}
   \mbox{\includegraphics[width=7cm,height=5cm]{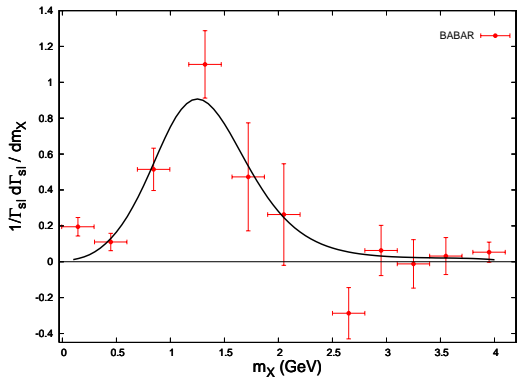}}
 \end{tabular}
\end{center}
\caption{\small{Comparison between theory and Babar data for
invariant hadron mass distribution in semileptonic decays for
$\alpha_S(m_Z) \, = \, 0.119$.}}
 \label{slmxbab}
\end{figure}

Among semileptonic decays, let us consider the invariant hadron
mass distribution, as shown in Fig.~\ref{slmxbab}, in comparison
with data from the BaBar collaboration \cite{Aubert:2006qi}.
Points with $m_X \, < \, 400$ MeV, which are dominated by the
$\pi$ peak, are discarded, as well as  points with $m_X \, > \,
2.6$ GeV, which give basically no information on the signal. Once
again, there is a good agreement between theory and data.

\begin{figure}[htbp]
\begin{center}
 \begin{tabular}{c}
   \mbox{\includegraphics[width=7cm,height=5cm]{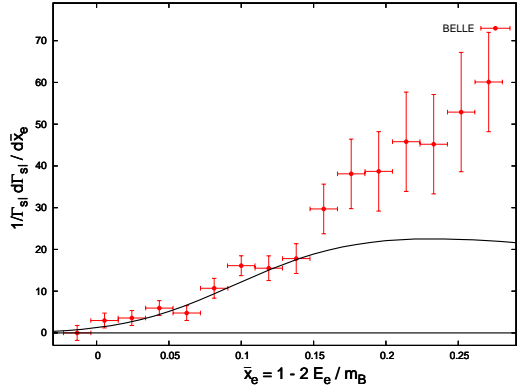}}
 \end{tabular}
\end{center}
\caption{\small{Comparison between theory and Belle data for the
electron spectrum in semileptonic decay for $\alpha_S(m_Z) \, = \,
0.135$. The data and the theory are normalized to one in the charm
background free region $0 \, < \, \bar{x}_e \, < \, 0.125$.}}
 \label{eebelle}
\end{figure}

The only  distribution where the agreement with data is less
satisfying is the electron spectrum in semi-leptonic decays, as
shown in Fig.~\ref{eebelle}, where data from Belle collaboration
have been used \cite{belleel}. Theory  predicts a harder spectrum,
with a broad maximum around $2.1$ GeV, while data peak at lower
energies. More accurate data, and possibly perturbative
calculations in the complete  NNLO approximation, are needed to
clarify such discrepancy.

%%%%%%%%%%%%%%%%%%%%%%%%%%%%%%%%%%%%%%%%%%%%%%%%%%%%%%%%%%%%%%%%%%%%%%%%%%%%%%%%%%%%%%%%%%%%

\end{document}